\documentclass[floatfix,showpacs,showkeys,amsmath,preprintnumbers,nofootinbib,onecolumn] {revtex4}
\usepackage{graphicx}
\usepackage{colordvi}
\usepackage{amsmath}
\usepackage{amssymb}
\usepackage{latexsym}
\usepackage{hyperref}
    \def\be{\begin{equation}}
    \def\ee{\end{equation}}
    \def\bea{\begin{eqnarray}}
    \def\eea{\end{eqnarray}}
\begin{document}
\title{Interacting Holographic Dark Energy, \\ the Present Accelerated Expansion and Black Holes}
\author{Bibekananda Nayak}
\affiliation{{P.G. Department of Applied Physics and Ballistics, Fakir Mohan University, Balasore, Odisha-756019, India} \\
E-mail: {bibekanandafm@gmail.com}}
\begin{abstract}
We study the evolution of the universe by assuming an integrated model, which involves interacting dark energy and holographic principle with Hubble scale as IR cutoff. First we determined the interaction rate at which matter is converting to dark energy. In the next step, we evaluated the equation of state parameter which describes the nature of dark energy. Our result predicts that the present state of the universe is dominated by quintessence type dark energy and it will become phantom dominated in near future. Again our analysis successfully addresses the problem of present accelerated expansion of the universe and softens the coincidence problem.  We also found that the universe was previously undergoing a decelerated phase of expansion and transition from deceleration to acceleration would occur at a time $t_{q=0}=0.732 t_0$, where $t_0$ is the present age of the universe. Finally, we discuss the evolution of Black Holes in this environment.
\end{abstract}
\maketitle
\section{Introduction}
One of the most intriguing discovery of modern day cosmology is the recent accelerated expansion of the universe, which was first predicted by the observations of Supernova Ia \cite{1,2} and subsequently confirmed by the observations of Cosmic Microwave Background Radiation \cite{3,4}, Weak Lensing \cite{5}, Large-scale Structure \cite{6,7,8}, WMAP Data \cite{9} and Planck Data \cite{10}. As a possible theoretical explanation of this major break though, it is considered that the most part of the present universe is made of a form of energy that exerts a negative pressure and drives the acceleration. The unknown nature of such energy brings it a name ‘Dark Energy’ and recent observational data \cite{10} suggests that 68.3 percentage of the present universe is filled with this unknown form of energy. On the other hand, the rest contents of the universe is gravitating matter with a large part is non-baryonic and is called ‘Dark matter’ again due to its nature.   

Due to lack of any concrete knowledge about the nature of the dark energy, there are a number of proposed candidates and their number is increasing day by day. Among them the natural and simplest candidate is cosmological constant. But it suffers from fine-tuning problem: mismatching in the magnitude of cosmological constant as predicted by field theory and present observation by 123 order. So dynamical dark energy models such as Quintessence \cite{11,12,13} and K-essence \cite{14} are proposed. Again there are discussion on an exotic form of dark energy, named as Phantom energy \cite{15,16,uv}, which violates dominant energy condition. There exist also ‘modified matter’ dynamical dark energy models like Chaplygin gas \cite{17,18} and also many modified theory of gravity like $f(R)$ gravity \cite{19,20}. But most of them are artificially constructed in the sense that it introduces too many free parameters to able to fit with observational data or not able to explain all features of the universe, like for example coincidence problem: why the observed values of the cold dark matter density and dark energy density are of the same order of magnitude today although they differently evolve during the expansion of the universe.

A new alternative to the solution of dark energy problem may be found in the Holographic Principle \cite{21,22,23}. According to the Holographic principle, the number of degrees of freedom in a bound system should scales with its boundary area not with its volume. By applying this principle to Cosmology, Cohen et al. \cite{24} found an upper bound on the entropy contained of the universe. For a system with size L and Ultra Violet (UV) cut-off Λ without decaying into black hole, it is required that the total energy in a region of size L should not exceed the mass of a black hole of the same size, i.e. $L^3\rho_\Lambda \leq LM_{pl}^2$. The largest allowed L is the one that saturating this inequality, so $\rho_\Lambda=3c^2M_{pl}^2L^{-2}$, where c is a constant, $\rho_\Lambda$ is the quantum zero point energy density and $M_{pl}$ is the Planck mass. It just means a duality between UV cut-off and Infrared (IR) cut-off, where UV cut-off is related to vacuum energy and IR cut-off is related to the large scale of the universe. In literature \cite{25,26,27,28,29,30,31,32}, it is considered that this holographic dark energy interacts with matter during the evolution of the Universe with Hubble scale, particle horizon or event horizon as IR cut-off.

Since dark energy and dark matter comprise approximately 96 percentage of the total energy density of the universe with unknown character and origin, sometimes it is assumed that perhaps the dark energy
and dark matter are coupled to each other so that they behave like a single dark fluid. Although this consideration sounds slightly phenomenological but this possibility cannot be ruled out by any observations. So, one can of course think of some interaction between these two fields. In fact, the standard cosmological laws can be retrieved at any time under the no interaction limit. Additionally, the dynamics of the universe in presence of any coupling between dark energy and dark matter becomes quite richer with many possibilities. Again from the particle physics point of view, any two fields can interact with each other. Since both dark energy and dark matter can be thought to be of some fields, for instance some scalar field, hence the idea behind the dark energy-dark matter interaction has a strong support from the particle physics side. The idea of coupling in the dark sectors was initiated by Wetterich \cite{wetterich} and subsequently discussed by Amendola \cite{amendola} and others. So far, this interacting scenario has been explored in the context of current cosmology with some interesting outcomes. Particularly, the coupling between the dark energy and the dark matter may provide an explanation to the coincidence problem \cite{ccp}, a generic problem in the dynamical dark energy models and even in the $\Lambda$-Cold Dark Matter ($\Lambda$CDM) cosmology. In the last couple of years, a rigorous analysis has been performed in the field of interacting dark energy by several authors with many interesting possibilities, see for instance \cite{a1,a2,a3,a4,a5,a6,a7,a8,a9,a10,a11,a12}.

In the present work, we use an interacting holographic dark energy model with Hubble scale as IR cut-off and study the evolution of the universe. First we calculate the equation of state parameter of the dark energy and then find deceleration parameter. We next evaluate the transition time from decelerated to present accelerated expansion.  We found that our analysis successfully addresses the problem of present accelerated expansion of the universe and coincidence problem. Also it predicts that present universe is dominated by quintessence type dark energy. Finally, we discuss the evolution of black holes in this interacting dark universe.
\section{Interacting Dark energy model}
For a spatially flat ($k=0$) FRW universe with scale factor ‘$a$’ and filled with dust and dark energy, the Friedman equations take the form
\bea \label{1}
\Big(\frac{\dot{a}}{a}\Big)^2 =\Big(\frac{8\pi G}{3}\Big)(\rho_{m}+\rho_{x}),
\eea       
\bea \label{2}
2\Big(\frac{\ddot{a}}{a}\Big)+ \Big(\frac{\dot{a}}{a}\Big)^2=-8\pi G p_{x}
\eea                                                                               
and the energy conservation equation becomes
\bea \label{3}
(\dot{\rho_x}+\dot{\rho_m})+3H(\rho_m+\rho_x+p_x)=0
\eea                                            
where $H={\frac{\dot{a}}{a}}$ is the Hubble parameter,
      $\rho_{x}$ = dark energy density,
      $\rho_{m}$ = matter energy density,
and   $p_x$ = pressure of the dark energy.

Now we use a dark energy model \cite{34} which rests on following three assumptions:\\
(i) The dark energy density is derived using holographic principle and is given by 
\bea \label{4}
\rho_x = \frac{3c^2}{8\pi G}L^{-2},
\eea  
where c is a dimensionless constant of O(1), L is IR cut-off and ${M_{pl}}^2 \approx \frac{1}{8\pi G}$.\\
(ii) IR cutoff is taken as the inverse of Hubble scale, i.e. $L=H^{-1}$.
So we can write
\bea \label{5}
\rho_x = \frac{3c^2}{8 \pi G}H^2.
\eea  
(iii) Matter and dark energy do not conserve separately but they interact with each other and one may grow at the expense of the other.

So the energy conservation equation in the presence of dark energy can be written as
\bea \label{6}
\dot{\rho_m}+3H\rho_m=Q,\nonumber \\ \dot{\rho_x}+3H(1+\omega)\rho_x=-Q.
\eea
Where $Q=\Gamma\rho_x$ with $\Gamma>0$ is the interaction rate having dimension of Hubble parameter and $\omega={\frac{p_x}{\rho_x}}$ denotes the equation of state parameter for the dark energy.
Now equation (\ref{6}) can be written as
\bea \label{7}
\dot{\rho_x}=-\{\Gamma+3H(1+\omega)\}\rho_x.
\eea  
Again taking derivative of equation (\ref{5}) with respect to time, we get
\bea \label{8}
\dot{\rho_x}=2 \rho_x \frac{\dot{H}}{H}.
\eea

In our model, we assume that the universe started with only dust and dark energy appeared due to its decay with $\Gamma$ as the interaction rate. By considering observational data \cite{10} that at present $68.3$ percentage of the universe is filled with dark energy and the rest are matter which has been achieved in $13.82\times 10^9$ years through their interaction, one can estimate that
\bea \label{9}
\Gamma\approx 4.942  \times 10^{-11}(yr)^{-1}.
\eea
This, further, implies that at any time t, the ratio of dark matter density to dark energy density should be given as 
\bea \label{10}
r=\frac{\rho_m}{\rho_x}=\frac{1-\Gamma t}{\Gamma t}.
\eea
The variation of $r$ with time is shown in Figure-1.

\begin{figure}[h]
\centering
\includegraphics[scale=0.8]{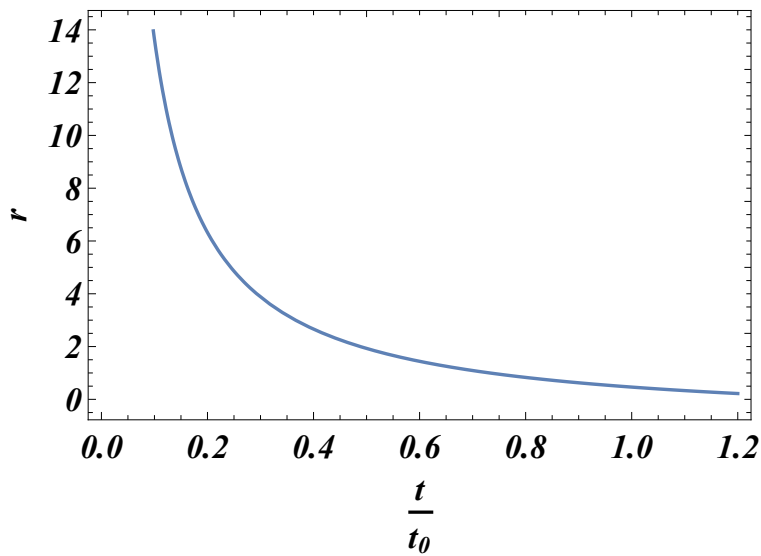}
\caption{Variation of $r=\frac{\rho_m}{\rho_x}$ with time}
\end{figure}

\section{Equation of state parameter of dark energy and Deceleration parameter}
According to our model, the universe is filled with matter and dark energy. But standard model of cosmology, where dark energy is absent, assumes that the universe was radiation dominated in the early time and it becomes matter dominated later and in both the era, scale factor obeys power law cosmology. Extending this power law cosmology to dark energy dominated era, here, we discuss the evolution of the universe in three different eras separately.

By comparing equation (\ref{7}) with equation (\ref{8}), one can find an expression for equation of state parameter of dark energy as
\bea \label{11}
\omega=-1-\frac{\Gamma}{3H}-\frac{2}{3}\frac{\dot{H}}{H^2}.
\eea

But  for general power law cosmology, scale factor varies like $a(t) \propto t^{\beta}$ and hence the equation of state parameter of dark energy becomes
\bea \label{12}
\omega=-1-\frac{\Gamma t}{3\beta}+\frac{2}{3\beta} .
\eea

Again Hubble’s law explains the expansion of the universe but whether the expansion is accelerating one or decelerating one, it can be determined by deceleration parameter. The deceleration parameter is defined as
\bea \label{13}
q={\frac{-{\ddot{a}}a}{{\dot{a}}^2}}.
\eea
From equation (\ref{2}), one can get
\bea \label{14}
2q-1=\frac{8\pi G p_x}{H^2}
\eea
which on simplification gives
\bea \label{15}
q={\frac{1}{2}}+{\frac{3}{2}}{\frac{1}{(1+r)}}{\omega}.
\eea
Now using equations (\ref{10}) and (\ref{12}), we get
\bea \label{16}
q={\frac{1}{2}}+{\frac{3}{2}} \Gamma t \Big[-1-\frac{\Gamma t}{3\beta}+\frac{2}{3\beta}\Big].
\eea
In radiation dominated era, $\beta=\frac{1}{2}$. So
\bea \label{17}
\omega=\frac{1}{3}-{\frac{2}{3}}\Gamma t.
\eea
and 
\bea \label{18}
q={\frac{1}{2}}+{\frac{1}{2}} \Gamma t (1-2 \Gamma t).
\eea
For matter dominated era, $\beta=\frac{2}{3}$. So
\bea \label{19}
\omega=-\frac{\Gamma t}{2}.
\eea
and
\bea \label{20}
q={\frac{1}{2}}-\frac{3}{4} (\Gamma t)^2 .
\eea

But for present era, power law cosmologies, high redshift data and present accelerated expansion of the universe demands $\beta$ to be greater than $1$ \cite{gad, zhz, avi} though exact value of $\beta$ has not been ascertained. Since present era is dark energy dominated, we construct the Table I by using equations (\ref{12}) and (\ref{16}), where subscript $0$ refers to present value. 

\begin{table}[ht]
\center
\caption{The present values of dark energy density parameter and deceleration parameter, and variation of transition times and transition redshifts from deceleration to acceleration for different values of $\beta$ are shown in the Table.}
\begin{tabular}{|c|c|c|c|c|}\hline
\hspace{0.1 in} $\beta$  \hspace {0.1 in}  &  \hspace {0.1 in} $\omega_0$  \hspace {0.1 in}    &  \hspace {0.1 in} $q_0$  \hspace {0.1 in}    & Transition Time & Transition Redshift \\
$ $$ $$ $$ $&  &  & $(t_{q=0})$ & $(z_{q=0})= (\frac{t_0}{t_{q=0}})^{3/2}-1$ \\ \hline 
1.1 & -0.601 & -0.116 & 0.885$t_0$ & 0.144\\
1.2  & -0.634 & -0.15 & 0.815$t_0$ & 0.278\\
1.3 & -0.662 & -0.179 & 0.782$t_0$ & 0.377\\
1.4 & -0.686 & -0.203 & 0.754$t_0$ & 0.485\\
1.5 & -0.707 & -0.225 & 0.732$t_0$ & 0.597\\
1.6 & -0.726 & -0.243 & 0.713$t_0$ & 0.718\\
1.7 & -0.742 & -0.26 & 0.696$t_0$ & 0.852\\
1.8 & -0.756 & -0.275 & 0.682$t_0$ & 0.992\\
1.9 & -0.769 & -0.288 & 0.669$t_0$ & 1.146\\
2.0 & -0.78  & -0.3 & 0.658$t_0$ & 1.309\\\hline
\end{tabular}
\end{table}

Comparing with the observational constraint on transition redshift as $z_{q=0}\approx 0.6$ \cite{yy, jras, mri}, we found $\beta$ to be $\frac{3}{2}$. This, in turn, determines the present value of equation of state parameter of dark energy and deceleration parameter as $\omega_0 \approx -0.707$  and $q_0 \approx -0.225$ respectively. Which indicates that the expansion of the universe is presently accelerating one and the universe is presently dominated by quintessence type dark energy. We also found that the transition time from deceleration to acceleration would be $t_{q=0} \approx 0.732 t_0$ which is in agreement with observation \cite{35}. Again from equation (\ref{12}), one can find that the universe will be phantom dominated in near future at time $t=2.928 t_0$. The evolution of equation of state parameter $\omega$ and deceleration parameter with time are shown in Figure-2 and Figure-3 respectively.

\begin{figure}[h]
\centering
\includegraphics[scale=0.8]{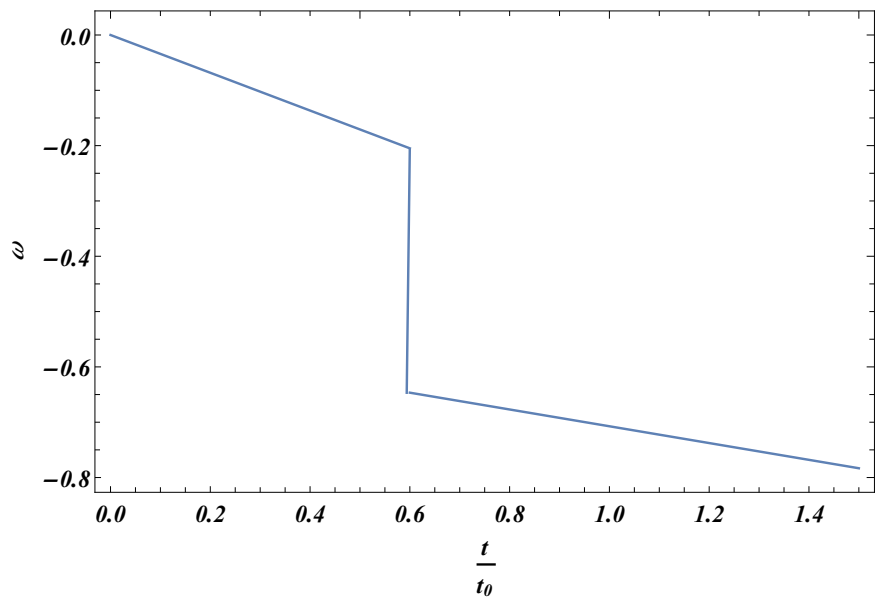}
\caption{Variation of equation of state parameter with time}
\end{figure}

\begin{figure}[h]
\centering
\includegraphics[scale=0.8]{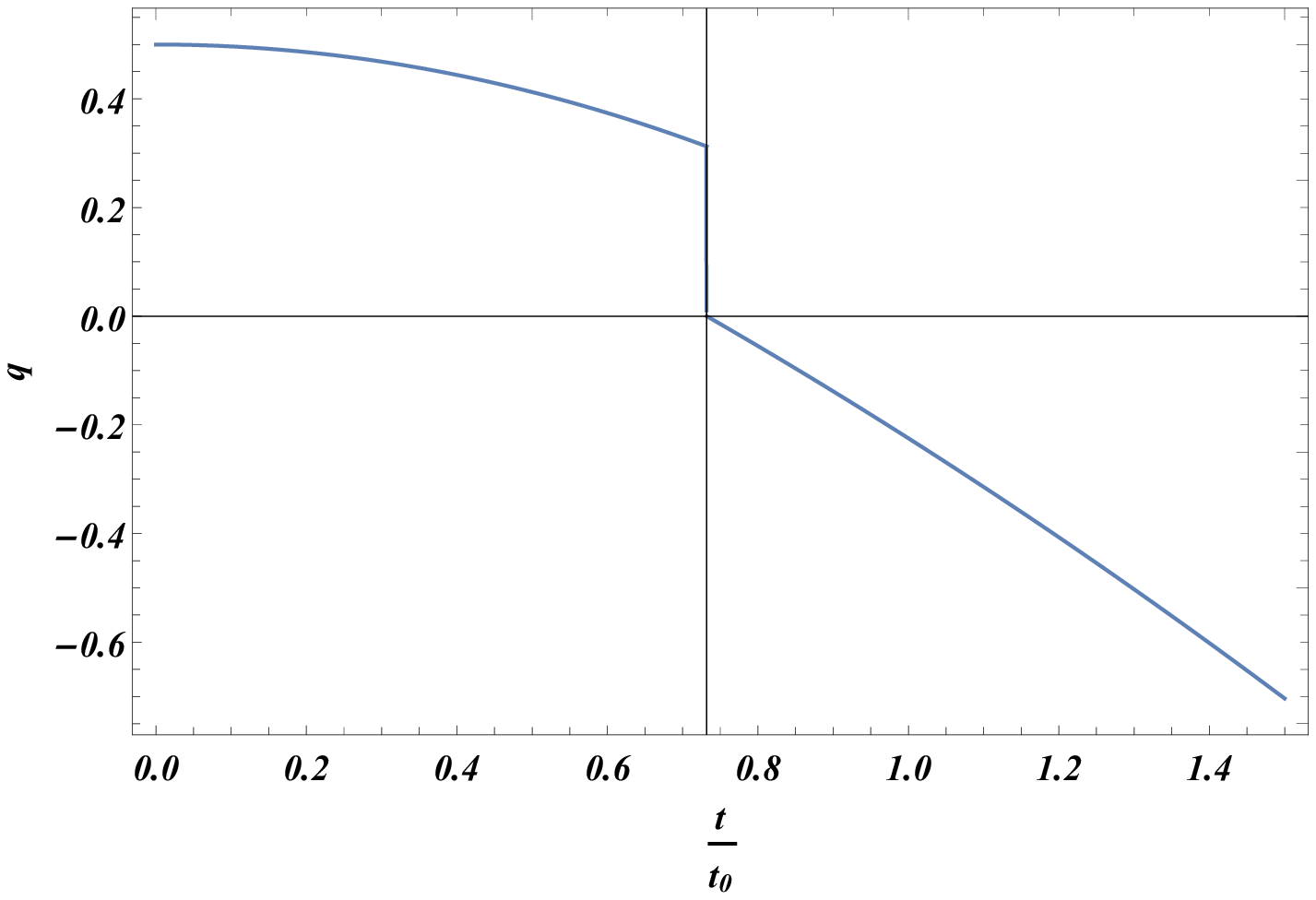}
\caption{Variation of deceleration parameter with time}
\end{figure}

\section{Coincidence Problem}
The current observations indicate that the densities of dark energy and dark matter are of the same order of magnitude today. This seems to imply that we are currently living in a very special period of the cosmic history, because dark energy and dark matter evolve differently during expansion of the universe. Within the context of Standard model, a density ratio of the order of one just at the present epoch can be seen as coincidence since it requires very special initial conditions in the early universe. The corresponding `why now' question constitutes the cosmological ``coincidence problem". But in very early universe and in the far-future universe, the energy densities differ by many orders of magnitude. Since $r$ is the ratio of energy densities, mathematically $\frac{\dot{r}}{r}$ is required to be very small for present epoch.

Now we calculate $\dot{r}$ from the definition of $r=\frac{\rho_m}{\rho_x}$ as
\bea \label{21}
\dot{r}=r\Big(\frac{\dot\rho_m}{\rho_m}-\frac{\dot\rho_x}{\rho_x}\Big).
\eea
Using energy conservation equation (\ref{6}), we get
\bea \label{22}
\frac{\dot{r}}{r}=3H\Big[\omega+\frac{\Gamma}{3H}\Big(1+\frac{1}{r}\Big)\Big].
\eea
Putting current values of various parameters in the above equation (\ref{22}), we found
\bea \label{23}
\Big|\frac{\dot{r}}{r}\Big|_0=0.229 \times 3H_0.
\eea
Thus $r$ varies more slowly in this model than in the conventional $\Lambda$CDM model, where $\Big(\frac{\dot{r}}{r}\Big)_0=3H_0$.\\
The variation of coincidence parameter $\Big|\frac{\dot{r}}{r}\Big|$ with time is shown in Figure-4.
\begin{figure}[h]
\centering
\includegraphics[scale=0.6]{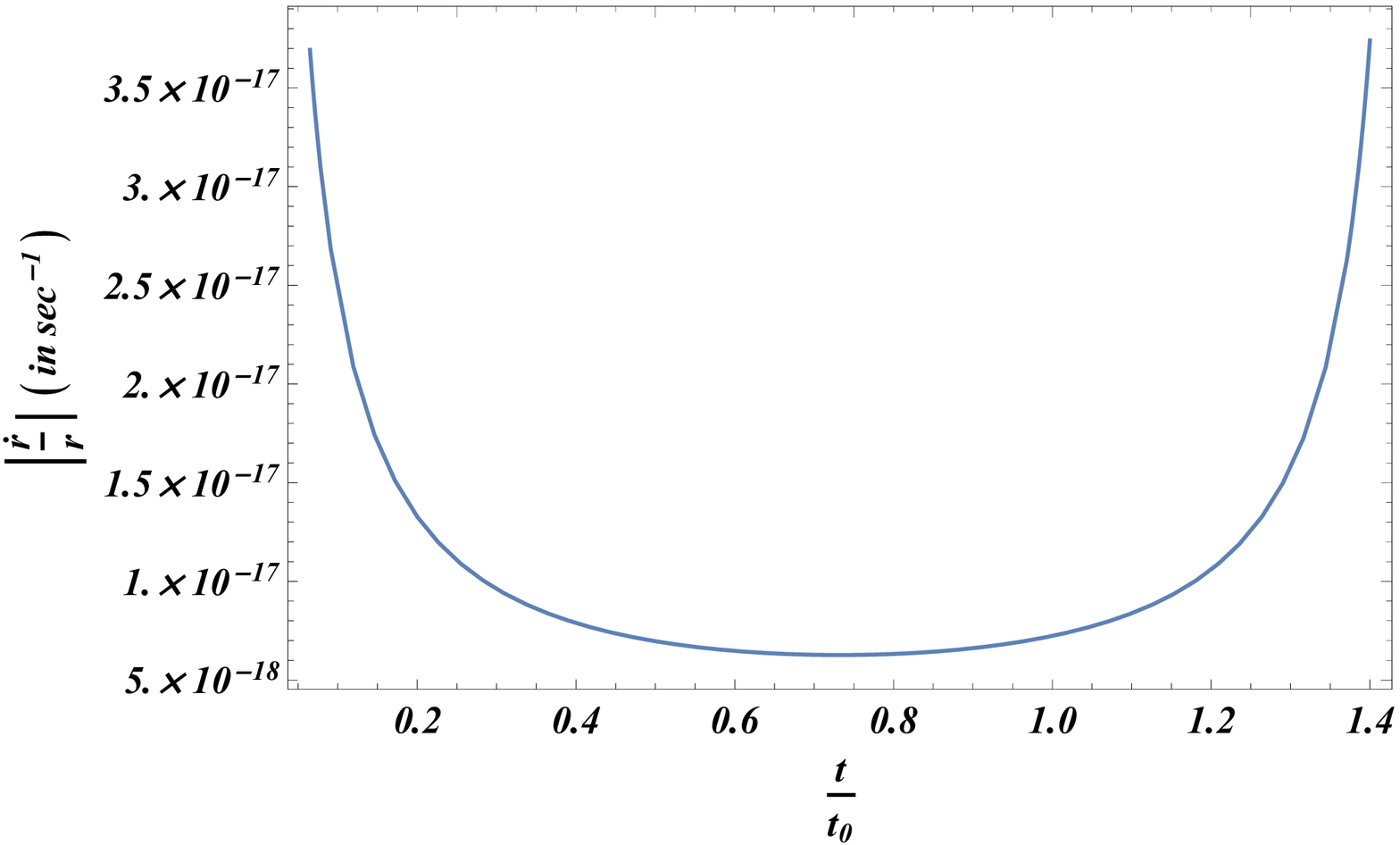}
\caption{Variation of $\Big|\frac{\dot{r}}{r}\Big|$ with time }
\end{figure}

\section{ Evolution of Black Holes in this environment}
Black hole is a region in space-time, where, gravitational field is so strong that even light can not escape from it.
In the usual formation scenarios, the typical mass of a black hole at the formation could be as large as the
mass contained in the Hubble volume $M_H$ ranging down to about
$10^{-4}M_H$ \cite{36}. Black holes can thus span enormous mass range starting from Planck mass to
few order of solar mass. In general, it is considered that all the black holes are formed by the time of matter-radiation equality $t_e$, 
which is assumed to be occurred when the universe is nearly $10^{11}$ sec old. So the maximum formation mass of the black hole would be  $(M_H)_{t_e}=G^{-1}t_e \sim 10^{49}$ gm.
Again the formation masses of some of the black holes could be small enough to have evaporated
completely by the present epoch due to Hawking evaporation \cite{37}.  
Early evaporating black holes
could account for baryogenesis \cite{38,39,40} in the universe. On the other
hand, presently surviving  
black holes could act as seeds for structure formation and
could also form a significant component of
dark matter \cite{41,42,43,44,45}. 
Once formed, these black holes are affected both by Hawking evaporation and accretion: absorption of energy matter from the surroundings.
In literature so many works \cite{46,47,48,49,50} are found, involving absorption of radiation, matter and dark energy. We, here, discuss accretion of interacting dark energy as follows.

Babichev et al. \cite{16} have worked out a differential equation for explaining the accretion of phantom type dark energy by black holes. By generalizing their analysis to all kinds of dark energies, we found that the mass of a black hole can be changed due to presence of interacting dark energy as
\bea \label{24}
\dot{M}=4 \pi R^2_{bh} (\rho_x+p_x) 
\eea
where $R_{bh}=2GM$ is the radius of black hole.\\
On simplification, above equation (\ref{24}) gives
\bea \label{25}
\dot{M}=16\pi G^2 M^2 (1+\omega) \rho_x .
\eea
Using equations (\ref{1}), one can find
\bea \label{26}
\dot{M}=\frac{6G}{1+r} M^2 (1+\omega) H^2 .
\eea
Due to Hawking evaporation, the rate at which the mass of a black hole changes is given by \cite{51}
\bea \label{27}
\dot{M}= -\frac{a_H}{256 \pi^3} \frac{1}{G^2M^2}
\eea
where $a_H$ is the black body constant.\\
In interacting dark universe, thus, the evolution of black holes' mass is governed by the equation
\bea \label{28}
\dot{M}= -\frac{a_H}{256 \pi^3} \frac{1}{G^2M^2}+\frac{6G}{1+r} M^2 (1+\omega) H^2
\eea
Like deceleration parameter, here we consider three epochs separately.

\subsection{Radiation dominated era}
In radiation dominated era, the equation (\ref{26}) takes the form
\bea \label{29}
\dot{M}=\frac{6G}{1+r} M^2 (1+\omega) \frac{1}{4 t^2}.
\eea
After using equations (\ref{10}) and (\ref{17}), and performing simple calculations, equation (\ref{29}) can be written as
\bea \label{30}
\dot{M}=G\Big(\frac{M^2}{t}\Big)\Gamma (2-\Gamma t)
\eea
By solving above differential equation (\ref{30}), we get
\bea \label{31}
M=M_i\Big[1-2\Gamma t_i ln(\frac{t_i}{t})+ (\Gamma t_i)^2 \Big(\frac{t}{t_i}-1\Big) \Big]^{-1}
\eea 
where $M_i$ is the mass of black hole at formation time $t_i$.
Again here $\Gamma t_i \le 10^{-6}$, so equation (\ref{30}) gives $ M \approx M_i$. i.e. Mass of a black hole is not affected 
by the presence of interacting dark energy in radiation dominated era. In radiation dominated era, hence, only Hawking evaporation term contributes towards
evolution of the black holes.

\subsection{Matter dominated era}
Since environment is not suitable for black holes to be formed in matter-dominated era, we study the evolution of those black holes in matter and dark energy dominated era which are only formed during radiation dominated era. \\

In matter dominated era, the equation (\ref{26}) takes the form
\bea \label{32}
\dot{M}=\frac{6G}{1+r} M^2 (1+\omega) \frac{4}{9t^2}.
\eea
On simplification, above equation gives
\bea \label{33}
\dot{M}=\frac{8}{3}G\Gamma \frac{M^2}{t}\Big(1-\frac{\Gamma t}{2}\Big)
\eea
Now the evolution of black holes' mass in matter dominated era is governed by the equation
\bea \label{34}
\dot{M}= -\frac{a_H}{256 \pi^3} \frac{1}{G^2M^2}+\frac{8}{3}G\Gamma \frac{M^2}{t} \Big(1-\frac{\Gamma t}{2}\Big)
\eea

\subsection{Dark energy dominated era}
In dark energy dominated era, the equation (\ref{26}) takes the form
\bea \label{35}
\dot{M}=\frac{6G}{1+r} M^2 (1+\omega) \frac{9}{4t^2}.
\eea
On simplification, above equation gives
\bea \label{36}
\dot{M}=6 G\Gamma \frac{M^2}{t}\Big(1-\frac{\Gamma t}{2}\Big)
\eea
By solving equations (\ref{33}) and (\ref{36}) numerically, we plot the Figure-5 which shows
the variation of black holes' mass with time in the presence of interacting dark energy.

\begin{figure}[h]
\centering
\includegraphics[scale=0.6]{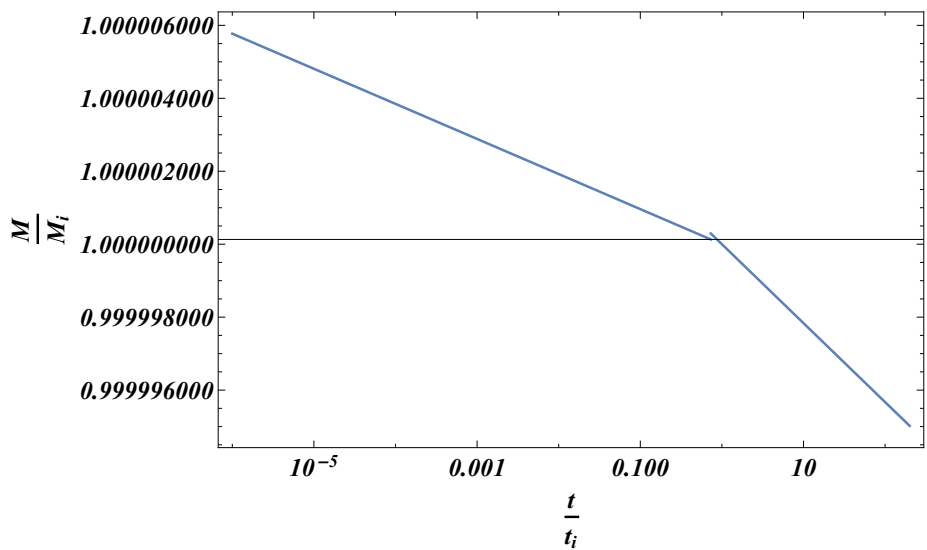}
\caption{Variation of black holes' mass with time having formation mass $10^{49}$ gm by considering only accretion of interacting dark energy.}
\end{figure}

From the Figure-5, we found that the mass of a black hole decreases due to accretion of interacting dark energy.

The evolution of black holes' mass in  dark energy dominated era is governed by the equation
\bea \label{37}
\dot{M}= -\frac{a_H}{256 \pi^3} \frac{1}{G^2M^2}+\frac{8}{3}G\Gamma \frac{M^2}{t} \Big(1-\frac{\Gamma t}{2}\Big)
\eea

Solving the equations (\ref{34}) and (\ref{37}) numerically and comparing with the numerical solution of equation (\ref{27}), we plot Figure-6 and
 construct Table-II, where the 2nd and 3rd columns give the evaporation times in absence and presence of interacting dark energy respectively and subscript $i$ refers to initial value.

\begin{figure}[h]
\centering
\includegraphics[scale=0.6]{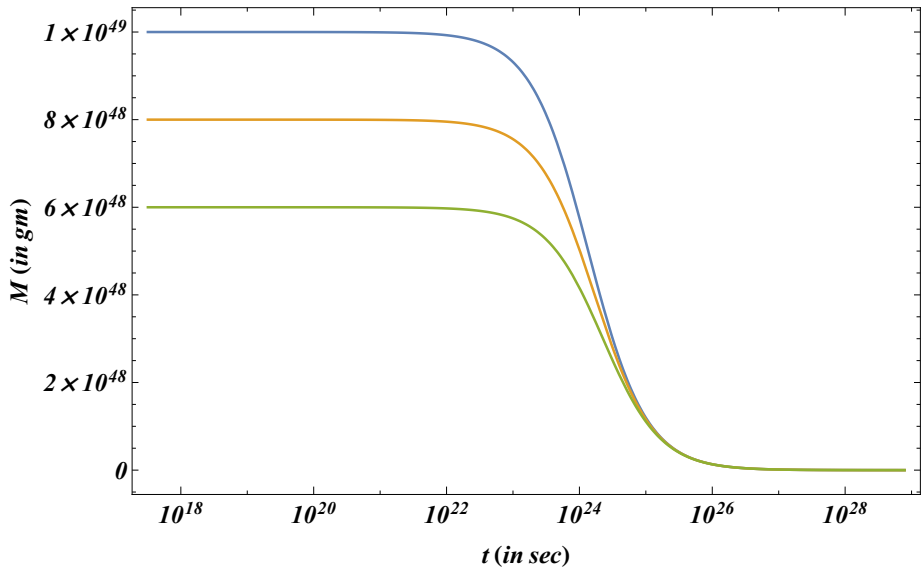}
\caption{Complete evolution of black holes in the presence of interacting dark energy is shown in the Figure}
\end{figure}

\begin{table}[ht]
\center
\caption{The  evaporation times of black holes in the presence of interacting dark energy }
\begin{tabular}{|c|c|c|c|}\hline
$t_i$ (in sec) & $M_i$ (in gm)  & $t_{evap}$ (in sec) & $(t_{evap})_{int-de}$ (in sec)\\\hline 
$10^{-23}$ & $10^{15}$ & $3.33 \times10^{16}$ & $3.33 \times10^{16}$\\
$10^{-18}$ & $10^{20}$ & $3.33 \times10^{31}$ & $3.33 \times10^{31}$\\
$10^{-13}$ & $10^{25}$ & $3.33 \times10^{46}$ & $3.23 \times10^{46}$\\
$10^{-10}$ & $10^{28}$ & $3.33 \times10^{55}$ & $7.85 \times10^{47}$\\
$10^{-8}$  & $10^{30}$ & $3.33 \times10^{61}$ & $7.86 \times10^{47}$\\
$10^{-3}$  & $10^{35}$ & $3.33 \times10^{76}$ & $7.86 \times10^{47}$\\
$10^{2}$   & $10^{40}$ & $3.33 \times10^{91}$ & $7.86 \times10^{47}$\\
$10^{7}$   & $10^{45}$ & $3.33 \times10^{106}$ & $7.86 \times10^{47}$\\
$10^{11}$  & $10^{49}$ & $3.33 \times10^{118}$ & $7.86 \times10^{47}$\\\hline
\end{tabular}
\end{table}

From Table-II, it is clear that the evaporation of black holes become quicker in the presence of interacting dark energy. Particularly, the black holes whose formation mass is greater than $10^{28}$ gm, they will suddenly vanish at $7.86 \times10^{47}$ sec indicating a strong phantom domination \cite{epjc} by that time. But presently evaporating black holes are not affected by the presence of dark energy, so all observed astrophysical constraints on black holes would not be disturbed.

\section{Discussion and Conclusion}
In this study, we use interacting holographic dark energy model, where we take Hubble scale as IR cutoff. We assume that during the evolution of the universe, dark energy is created at the cost of matter. As a success of our work, we first determined the interaction rate at which matter is converting to dark energy. Then we calculated the equation of state parameter which describes the nature of dark energy. Our result predicts that the present state of the universe is dominated by quintessence type dark energy and it will be phantom dominated in near future at a time $t=2.928 t_0$. Again our model is successful in explaining the present accelerated expansion of the universe. Our result tells that the universe was previously undergoing a decelerated phase of expansion and transition from deceleration to acceleration would occur at $t_{q=0}=0.732 t_0$, where $t_0$ is the present age of the universe. So in this case, the transition is hastened in comparison with previous result \cite{34} where scalar-tensor theory is used. Our analysis also considerably softens the coincidence problem. Finally, we took a look on evolution of black holes in this environment. From our study, we found that the black holes, particularly formed after $10^{-13}$ sec would be affected by the presence of interacting dark energy. During their evolution, those black holes would loss their mass due to interacting dark energy and hence their evaporation would be quicker in comparison with standard scenario. Again all the black holes whose formation mass is greater than $10^{28}$ gm, will be evaporated at a particular time $t=7.86\times10^{47}$ sec. This can be explained by the fact that by that time the universe will become strongly phantom dominated \cite{epjc}. But the presence of interacting dark energy could not affect the presently evaporating black holes whose formation mass ($M_i$) is of the order of $ 10^{15}$ gm and thus all observed astrophysical constraints on black holes remain unaltered.

Thus our integrated model involving interacting dark energy and holographic principle with Hubble scale as IR cutoff can accommodate present accelerated expansion of the universe. It is also successful in determining the interaction rate between dark energy and matter, and the transition time from decelerated to accelerated expansion. Again it predicts that the present universe is dominated by quintessence type dark energy and softens the coincidence problem. Moreover, we found that black hole dynamics is strongly affected by the presence of interacting dark energy.
\\
\\
\noindent{\bf{Acknowledgement}}\\
This work is financially supported by UGC Start-Up-Grant Project of Dr. Bibekananda Nayak having Letter No. F. 30-390/2017 (BSR) of University Grants Commission, New Delhi. I am also thankful to Prof. L. P. Singh of Utkal University, Bhubaneswar for useful discussions.
\\

\end{document}